\documentclass{emulateapj}
\makeatletter
\begin{document}
\title{The Largest Gravitational Lens: MACS J0717.5+3745 ($\lowercase{z}$=0.546)}
%%%%%%%%%%%%%%%%%%%%%%%%%%%%%%%%%%%%%%%%%%%%%%%%%%%%%
\def\HI{\hbox{H~$\scriptstyle\rm I\ $}}
\def\arcs{$''$}
\def\arcm{$'$}
\def\sfrd{\,{\rm M_\odot\,yr^{-1}\,Mpc^{-3}}}
\def\emunits{\,{\rm ergs\,s^{-1}\,Hz^{-1}\,Mpc^{-3}}}
\def\be{\begin{equation}}
\def\ee{\end{equation}}
%%%%%%%%%%%%%%%%%%%%%%%%%%%%%%%%%%%%%%%%%%%%%%%%%%%%%
%\usepackage{amssymb}
%%\usepackage{natbib}
%\usepackage{alltt}
%\usepackage{epsfig}
%\usepackage{subfigure}
%%\usepackage{rotating}
%%\usepackage{supertabular}
%\usepackage{lscape}
%\usepackage{longtable}
%%\usepackage{aastex_hack}
%\usepackage{caption2}
%\usepackage{textcomp}
%\bibliographystyle{naturemag}

\author{Adi Zitrin, Tom Broadhurst, Yoel Rephaeli \& Sharon Sadeh}
\affil{School of Physics and Astronomy, Tel Aviv University, Israel}

%\altaffiltext{1}{School of Physics and Astronomy, Tel Aviv University, Israel}
\shorttitle{The Largest Gravitational Lens: MACS J0717.5+3745}
\shortauthors{Zitrin et al.}

\slugcomment{Accepted to the Astrophysical Journal Letters}

\begin{abstract}

We identify 13 sets of multiply-lensed galaxies around MACS
J0717.5+3745 ($z=0.546$), outlining a very large tangential critical
curve of major axis $\sim2.8\arcmin$, filling the field of HST/ACS.
The equivalent circular Einstein radius is $\theta_{e}= 55 \pm
3\arcsec$ (at an estimated source redshift of $z_{s}\sim2.5$),
corresponding to $r_e\simeq 350\pm 20~kpc$ at the cluster redshift, nearly
three times greater than that of A1689 ($r_e\simeq 140~kpc$ for $z_{s}=2.5$).
The mass enclosed by this critical curve is very large, $7.4\pm 0.5
\times 10^{14}M_{\odot}$ and only weakly model dependent, with a
relatively shallow mass profile within $r<250~kpc$, reflecting the
unrelaxed appearance of this cluster. This shallow profile generates
a much higher level of magnification than the well known relaxed
lensing clusters of higher concentration, so that the area of sky
exceeding a magnification of $>10\times$, is $\simeq 3.5\sq\arcmin$
for sources with $z\simeq 8$, making MACS~J0717.5+3745 a compelling
target for accessing faint objects at high redshift. We calculate that
only one such cluster, with $\theta_{e}\ge 55\arcsec$, is predicted
within $\sim 10^7$ Universes with $z\ge 0.55$, corresponding to a
virial mass $\ge 3\times 10^{15}\,M_{\odot}$, for the standard
$\Lambda CDM$ (WMAP5 parameters with $2\sigma$ uncertainties).

\end{abstract}
\section{Introduction}

Improvements in gravitational-lensing modeling together with the fabulous
image quality of the Advanced Camera have led to the identification of
increasing numbers of multiply-lensed images and the discovery of some
surprisingly large lenses (e.g., Kneib et al., 1996, Broadhurst et
al. 2005, Brada$\check{c}$ et al., 2008, Halkola et al. 2008,
Liesenborgs et al., 2008, Limousin et al., 2008, Zitrin et al. 2009,
Zitrin \& Broadhurst 2009).

The largest bound structures to form hierarchically are likely to have collapsed recently and hence should be found relatively locally, at low redshifts. The
larger volume available with increasing distance means that in
practice we cannot expect to reside next to the most massive
cluster. The Coma cluster, $z=0.023$, is the most
massive ``local'' cluster with a reliable weak lensing based mass of
$M_{vir}\sim 0.65 \times 10^{15}M_{\odot}$ (Gavazzi et al., 2009), similar
to earlier dynamical estimates $\sim0.8\times 10^{15}M_{\odot}$ (The \&
White, 1986, Geller et al., 1999). Other more distant clusters are known
from lensing to be more massive, such as A2218 at $z=0.16$ (Kneib et al.,
1996) and A1689 at $z=0.18$, with $1.6\pm0.2 \times 10^{15}M_{\odot}$
(Broadhurst et al., 2005, Umetsu \& Broadhurst, 2008), superseded by the
most distant Abell cluster, A370 at $z=0.37$ and the most luminous X-ray
cluster RXJ1347, $z=0.45$, with masses reliably determined from weak
lensing distortion and magnification of $\simeq 2\times
10^{15}M_{\odot}$ (Broadhurst et al. 2008).

Strong lensing is not seen for low redshift clusters ($z<0.15$)
because the central projected mass densities do not exceed the
critical density required to generate a sizable Einstein radius, which
scales inversely with lens distance, diverging at low
redshift. Similarly, at high redshift as the lens distance approaches
that of the distant sources, thus the critical density is too great for
strong lensing. Geometrically, lensing is optimized at intermediate redshifts,
where for a given mass the critical density for lensing is minimal, but
this is partly offset by the late hierarchical growth of high-mass systems. This trade-off results in estimates
of the amplitude of strong lensing to favor the redshift range
$z=0.2-0.4$, for NFW-like mass profiles (Broadhurst \& Barkana, 2008,
Oguri \& Blandford, 2009). Seemingly, all rich and X-ray luminous
clusters at modest redshift are found to have many multiply-lensed
images when examined with sufficient depth and resolution, implying
that the mass profiles of clusters are in general sufficiently peaked
that the critical density for lensing is exceeded.

The Einstein radii of the most massive clusters seem to be larger than
predicted by the $\Lambda CDM$ model (Broadhurst \& Barkana, 2008,
Puchwein \& Hilbert, 2009), based on the ``Millenium'' simulation
(Springel et al., 2005). This discrepancy is empirically supported by
the surprisingly concentrated mass profiles measured for such
clusters, when combining the inner strong lensing with the outer weak
lensing signal (Gavazzi et al., 2003, Broadhurst et al., 2005 \& 2008,
Limousin et al., 2008, Donnarumma et al., 2009, Oguri et al., 2009,
Umetsu et al., 2009, Zitrin et al., 2009).

These largest lensing clusters have proven to be excellent targets for
accessing the faint early Universe due to their large magnification
consistently providing the highest redshift
galaxies (Ebbels et al., 1996, Franx et al., 1997, Frye \& Broadhurst,
1998, Bouwens et al., 2004, Kneib et al., 2004, Bradley et al., 2008,
Zheng et al., 2009, see also Broadhurst, Taylor \& Peacock 1995). The larger sizes of lensing images has provided
increased spatial detail and the large magnifications permits observations
of increased spectral resolution, leading to the discovery that
metal enriched material is typically outflowing from galaxies at $z>4$
(Franx et al. 1997, Frye \& Broadhurst, 1998, Frye, Broadhurst \& Ben\'itez, 2002).

With the goal of discovering high redshift galaxies and to better
define the mass profiles of galaxy clusters in general, we have
combined data for a sample of well studied clusters. In this process
we have uncovered the extreme lensing properties of MACS J0717.5+3745
($z$=0.546), a cluster originally identified in the highly complete
sample of the most X-ray luminous clusters in the Universe (Ebeling et
al., 2004, 2007). This cluster is thought to be amongst the most
massive clusters known, forming part of an intricate dark matter (DM) filamentary
structure of $\sim$ 4 Mpc (Ebeling et al., 2004, Ma et al., 2008), whose
complex X-ray emission indicates ongoing merging (Ma et al., 2009)
accompanied by the most powerful known radio halo (van Weeren et al.,
2009, Bonafede et al., 2009).

In \S2 we describe the observations; the lensing
analysis is described in \S3; our results are presented in \S4 and in \S5 we discuss and
conclude them. Throughout the paper we adopt the standard cosmology
($\Omega_{\rm m0}=0.3$, $\Omega_{\Lambda0}=0.7$,
$h=0.7$). Accordingly, one arcsecond corresponds to 6.42 $kpc/h_{70}$
at the redshift of this cluster. The reference center of our analysis
is fixed near the center of the ACS frame at: RA = 07:17:31.65, Dec =
+37:45:03.12 (J2000.0).

\section{Observations}

The very X-ray luminous cluster MACS J0717.5+3745, which is the denser
north-western region of the large-scale filament found by
Ebeling et al. (2004), was imaged in April 2004 and in October
2005, 2006, with the Wide Field Channel (WFC) of the ACS installed on
HST. Integration times of $\sim 4500s$ were obtained through each of
the F555W and the F814W filters in the April 2004 run. We retrieved these images from the
Hubble Legacy Archive, along with a subsequent exposure in the F606W
band (January 2005, integration time of 1980s) to form a composite
high-resolution 3-color image. Several obvious close pairs of
multiply-lensed galaxies and giant arcs are immediately visible
throughout the full frame, with which we begin our modeling process
described below.

\section{Lensing Analysis}

 We apply our well tested approach to lens modeling, which we have
 applied successfully to A1689, Cl0024, and MACS J1149.5+2223,
 uncovering large numbers of multiply-lensed images in several
 clusters imaged with HST/ACS (Broadhurst et al., 2005, Zitrin et al.,
 2009, Zitrin \& Broadhurst 2009). The full details of this approach
 can be found in these papers. Briefly, the basic assumption adopted
 is that mass approximately traces light, so that the photometry of
 the red cluster member galaxies is the starting point for our model.

 Cluster member galaxies are identified as lying close to the cluster
 sequence by the photometry provided in the Hubble Legacy Archive.  We
 approximate the large scale distribution of matter by assigning a
 power-law mass profile to each galaxy, the sum of which is then
 smoothed. The degree of smoothing and the index of the power-law are
 the main free parameters. A worthwhile improvement in fitting the
 location of the lensed images is generally found by expanding to
 first order the gravitational potential of the smooth component,
 equivalent to a coherent shear describing the overall matter ellipticity, where the direction of the shear and
 its amplitude are free, allowing for some flexibility in the relation
 between the distribution of DM and the distribution of galaxies, which
 cannot be expected to trace each other in detail. The total
 deflection field $\vec\alpha_T(\vec\theta)$, consists of the galaxy
 component, $\vec{\alpha}_{gal}(\vec\theta)$, scaled by a factor
 $K_{gal}$, the cluster DM component $\vec\alpha_{DM}(\vec\theta)$,
 scaled by (1-$K_{gal}$), and the external shear component
 $\vec\alpha_{ex}(\vec\theta)$:

\begin{equation}
\label{defTotAdd}
\vec\alpha_T(\vec\theta)= K_{gal} \vec{\alpha}_{gal}(\vec\theta)
+(1-K_{gal}) \vec\alpha_{DM}(\vec\theta)
+\vec\alpha_{ex}(\vec\theta),
\end{equation}
where the deflection field at position $\vec\theta_m$
due to the external shear,
$\vec{\alpha}_{ex}(\vec\theta_m)=(\alpha_{ex,x},\alpha_{ex,y})$,
is given by:
\begin{equation}
\label{shearsx}
\alpha_{ex,x}(\vec\theta_m)
= |\gamma| \cos(2\phi_{\gamma})\Delta x_m
+ |\gamma| \sin(2\phi_{\gamma})\Delta y_m,
\end{equation}
\begin{equation}
\label{shearsy}
\alpha_{ex,y}(\vec\theta_m)
= |\gamma| \sin(2\phi_{\gamma})\Delta x_m -
  |\gamma| \cos(2\phi_{\gamma})\Delta y_m,
\end{equation}

where $(\Delta x_m,\Delta y_m)$ is the displacement vector of the
position $\vec\theta_m$ with respect to a fiducial reference position,
which we take as the lower-left pixel position $(1,1)$, and
$\phi_{\gamma}$ is the position angle of the spin-2 external
gravitational shear measured anti-clockwise from the $x$-axis.

We lens candidate galaxies back to the source plane using the derived
deflection field, and then relens this source plane to predict the
detailed appearance and location of additional counter images, which
are then searched for in the data (see Figures 1 and 2). The
fit is assessed by the RMS uncertainty in the image plane:

\begin{equation} \label{RMS}
RMS_{images}^{2}=\sum_{i} ((x_{i}^{'}-x_{i})^2 + (y_{i}^{'}-y_{i})^2) ~/ ~N_{images},
\end{equation}
where $x_{i}^{'}$ and $y_{i}^{'}$ are the locations given by the
model, and $x_{i}$ and $y_{i}$ are the real images location, and the
sum is over all $N_{images}$ images.

Importantly, this image-plane minimization does not suffer from the
well known bias involved with source plane minimization which
biasses solutions towards high magnification and hence
correspondingly shallower profiles. The model is successively
refined as additional sets of multiple images are identified and
incorporated to improve the fit (Zitrin et al., 2009). In addition,
we find that a bright foreground elliptical galaxy (RA=07:17:37.16,
DEC=+37:44:22.54) is important to include in the model as it locally
affects lensed images in the eastern to central part of the ACS
frame, including their relative lensing distance.

% This
%influence was used to constrain the relative scaling of this galaxy
%with respect to the regular cluster members, via the image-plane RMS
%of the reproduced images closer to this object.

In the above process we uncovered and used 34 multiply-lensed images,
corresponding to 13 lensed background galaxies, to constrain the mass
distribution and profile of this cluster. A lensed pair of red
drop-out objects which are not seen in the F555W band at an estimated
redshift of $z\sim4$ (system number 5, see Figure 1), helps us pin down
accurately the normalization of the deflection field by adopting the
lensing distance for this redshift which is nearly independent of
cosmology (see also Broadhurst et al., 2005, Zitrin et al., 2009). The image-plane RMS of our final model is very good, 2.2 $\arcsec$ per image, whereas for comparison, 
in A1689 an RMS of 3.2 $\arcsec$ was achieved by Broadhurst et al. (2005) and for Cl0024 an RMS of 2.5 $\arcsec$ was achieved by Zitrin et al. (2009).

\begin{figure}
%%[h!]
    %\epsfig{figure=figure1.eps,width=15cm}
%\epsscale{0.95}
%%\plotone{4clusters.ps}
%\vspace{0.5cm}
 \begin{center}
  \includegraphics[clip,width=80mm]{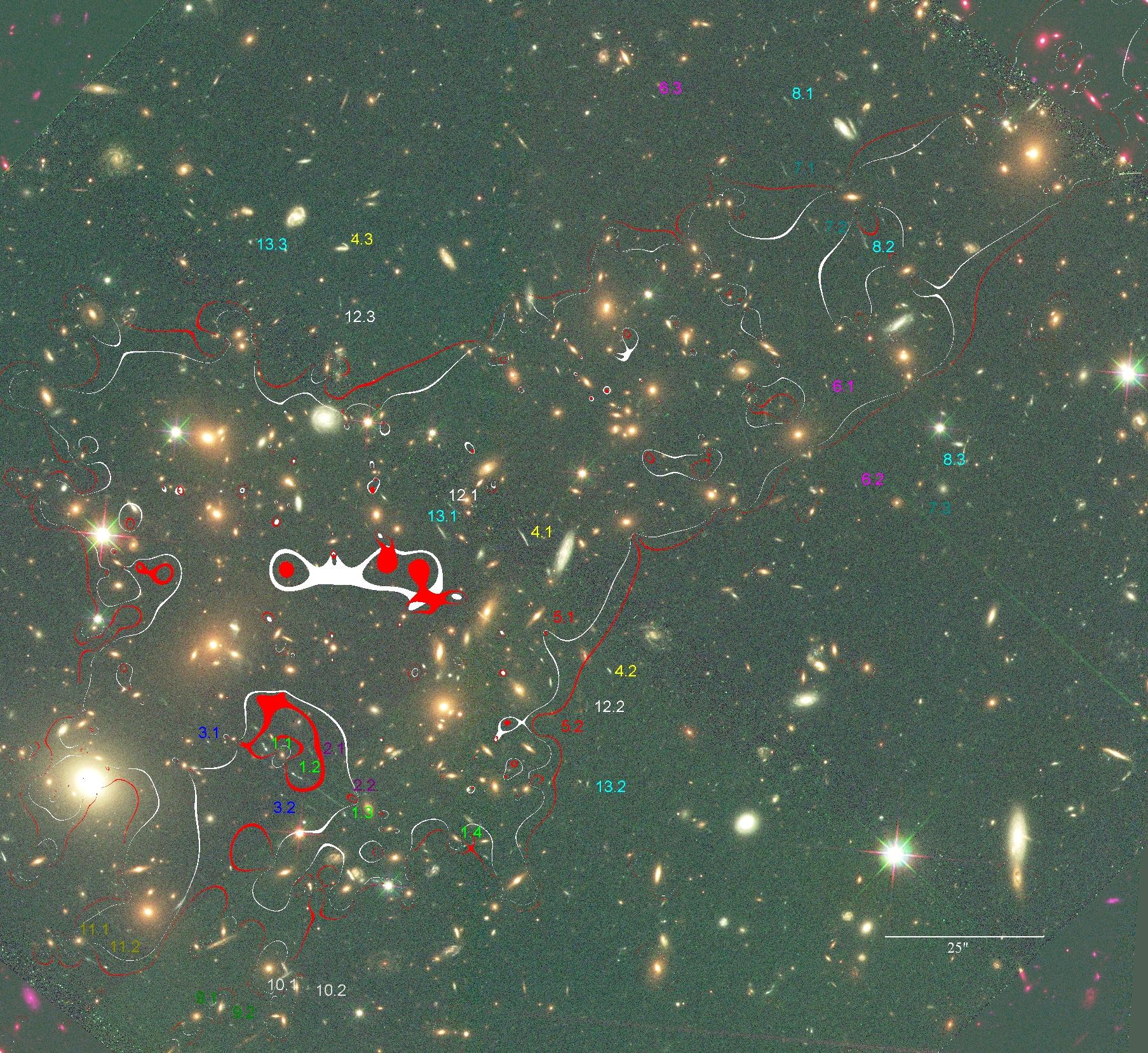}
 \end{center}
\caption{The cluster MACS J0717.5+3745 ($z=0.546$) imaged
with Hubble/ACS F555W, F606W, and F814W bands. The 34 multiply lensed
images identified by our model are numbered here. The white curve
overlaid shows the tangential critical curve corresponding to the
distance of system~1 at an estimated redshift of $z\sim2.5$, and which
passes through several close pairs of lensed images in this system. The larger
critical curve overlaid in red corresponds to the larger source
distance for the red dropout galaxy number ~5, at the estimated photometric
redshift of $z\sim4$. This large tangential critical curve encloses a
very large lensed region equivalent to $\sim400~kpc$ in radius at the
redshift of the cluster, $z=0.546$.}
\end{figure}

\begin{figure}
%%[h!]
    %\epsfig{figure=figure1.eps,width=15cm}
%\epsscale{0.95}
%%\plotone{4clusters.ps}
%\vspace{0.5cm}
 \begin{center}
  \includegraphics[clip,width=90mm]{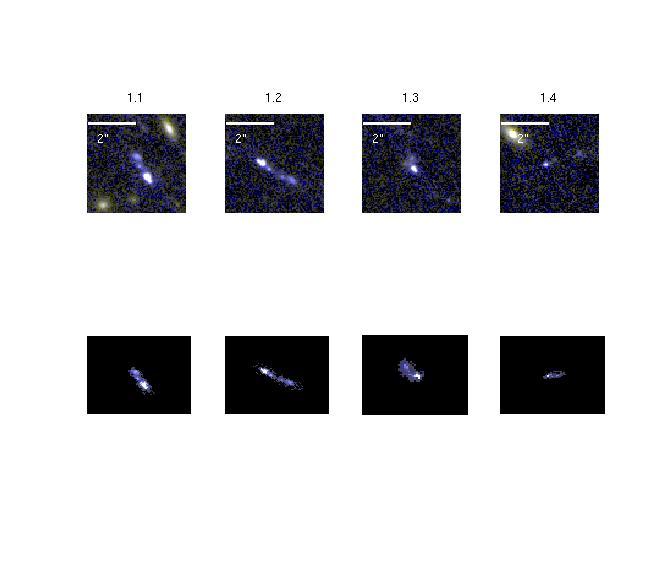}
 \end{center}
\caption{A reproduction of system~1 by our model. The
observed images are shown in the top row are compared with our model
generated images below. Each model image is generated using as input
the pixels of image 1.1 (except for the model image of 1.1 which is
generated by relensing the observed image 1.2) and delensing these
pixels back to the source plane and then relensing the source plane to
generate the counter images. It is clear that our model is successful
in demonstrating the multiply-lensed relation between
the four observed images.}
\end{figure}

\begin{figure}
%%[h!]
    %\epsfig{figure=figure1.eps,width=15cm}
%\epsscale{0.95}
%%\plotone{4clusters.ps}
%\vspace{0.5cm}
 \begin{center}
  \includegraphics[width=80mm]{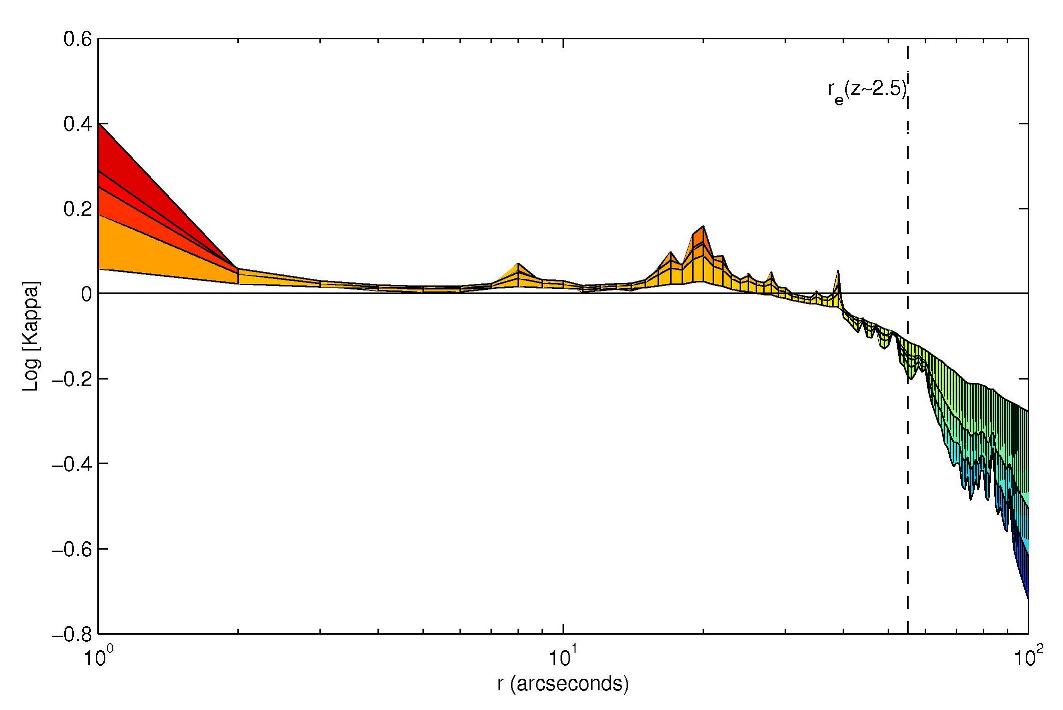}
 \end{center}
\caption{Radial surface mass density profile, $\kappa(r)$, in units of the
critical surface density, i.e.  $\kappa(r)=\Sigma(r)/\Sigma_{crit}$,
derived for the range of radius covered by all sets of multiple images
shown in Figure~2. The profile is shallow within $\sim 250~kpc$, at a
level close to the critical density for system~1 (black line). The vertical dashed line at
$\sim$55'' is the Einstein radius at the redshift of system~1. This profile was
measured circularly while centered on the central pixel of the 2D mass
distribution shown in Figure 4.}
\end{figure}

\begin{figure}
%%[h!]
    %\epsfig{figure=figure2.eps,width=15cm}
%\epsscale{0.95}
%%\plotone{4clusters.ps}
 \begin{center}
  \includegraphics[width=80mm]{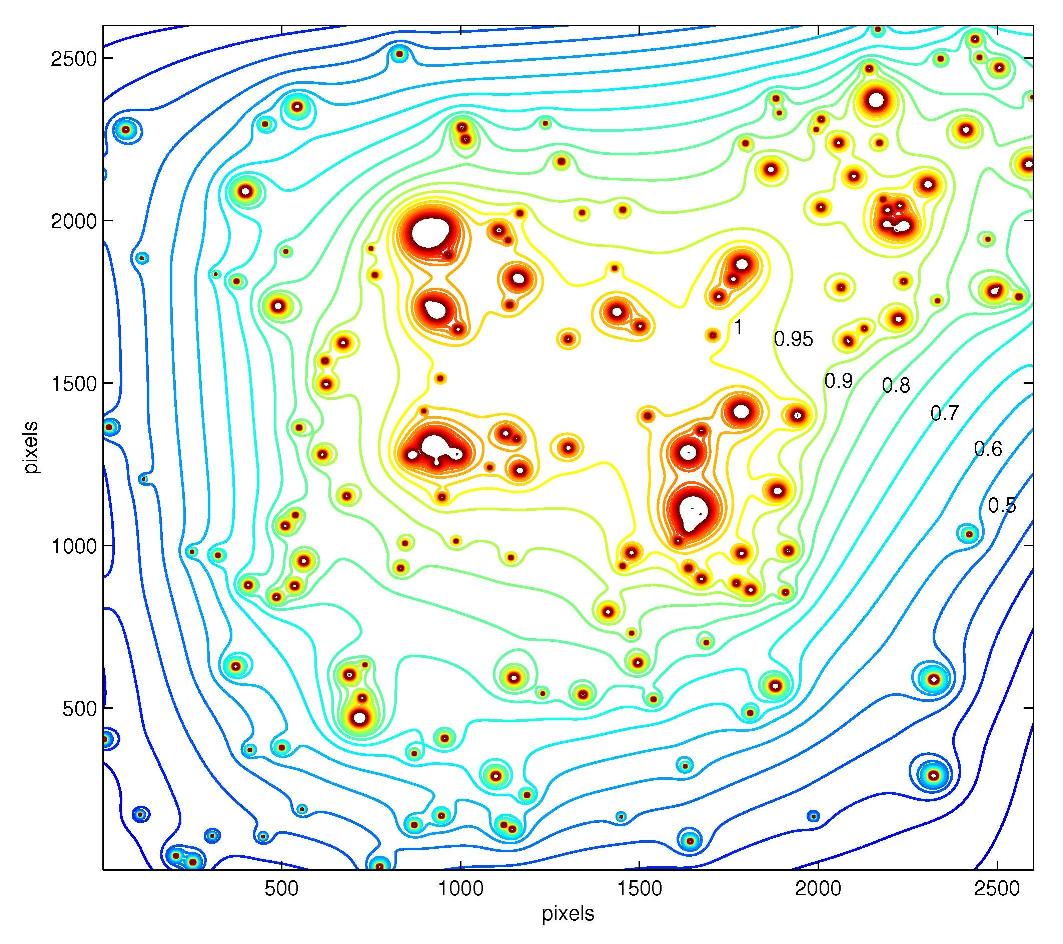}
 \end{center}
\caption{Zoomed-in 2D surface mass distribution ($\kappa$), in units of the
critical density. Contours are shown in linear units, derived from
the mass model constrained using 34 multiply-lensed images seen in
Figure 2. The axes are in ACS pixels. Note that the central mass distribution of
is rather flat reflecting the unrelaxed appearance
of this cluster.}
\end{figure}

\begin{figure}
%%[h!]
    %\epsfig{figure=figure2.eps,width=15cm}
%\epsscale{0.95}
%%\plotone{4clusters.ps}
 \begin{center}
  \includegraphics[width=80mm]{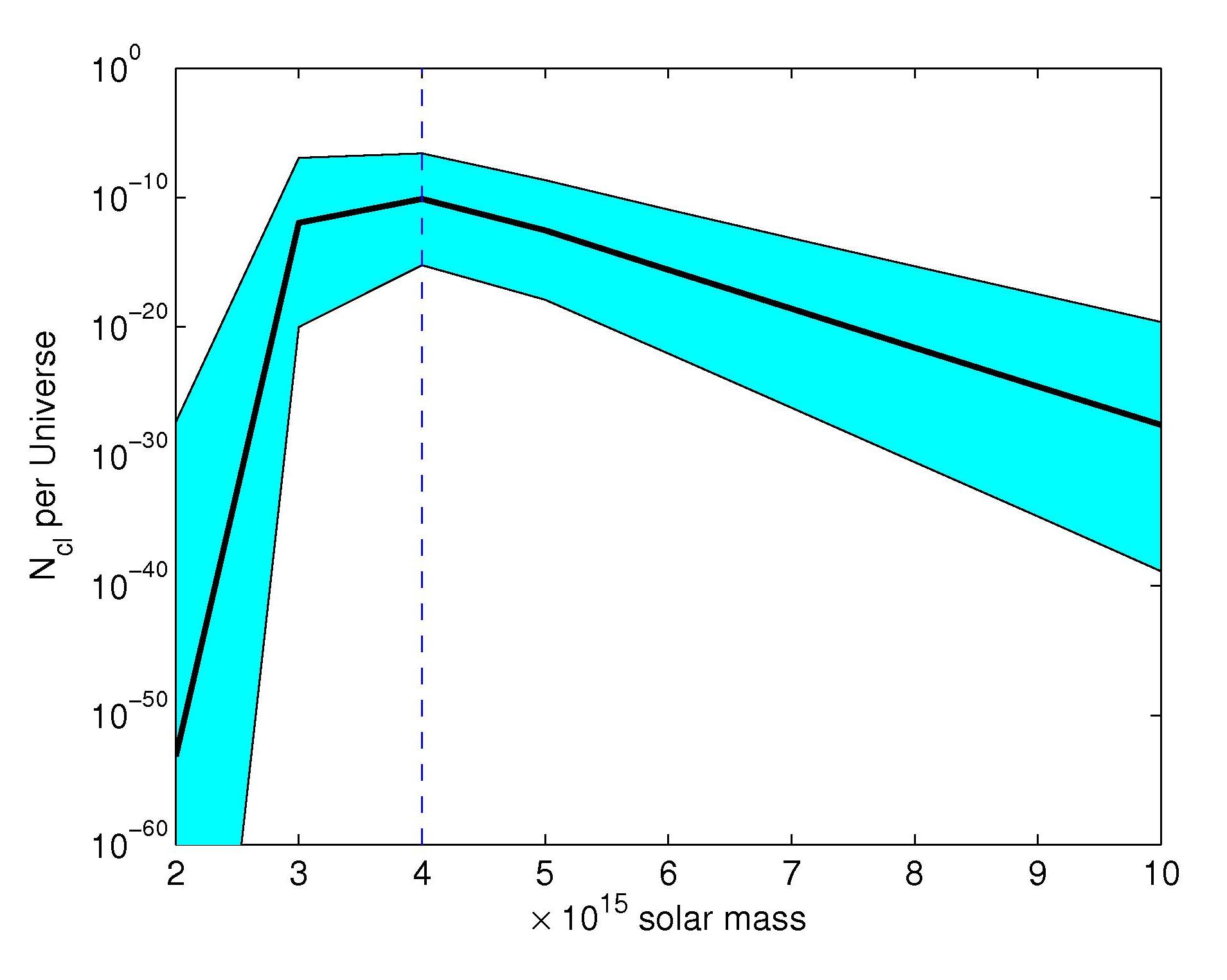}
 \end{center}
\caption{The number of clusters per universe with masses exceeding $M$, lying at
redshift $z\ge 0.55$, for which the predicted Einstein radius
is $\theta\ge 55\pm 3''$.  The shaded area reflects the variance
generated by incorporating into the the $\Lambda$CDM model the
$2\sigma$ uncertainty in the cosmological parameters $h$, $n$, and
$\sigma_8$, about the WMAP5 values, and also the $1\sigma$ uncertainty
in the measured $\theta_E$. This number peaks near $M\sim 4\times
10^{15}M_{\odot}$, with $N_{cl}\sim 2.7\cdot 10^{-7}$.}
\end{figure}

\vspace{1cm}
\section{Results}

The derived surface mass distribution is relatively shallow (see
Figure 3) in accordance with the unrelaxed appearance of this cluster,
which is known to be in the process of merging, with disturbed X-ray
emission, hot shocked regions (Ma et al., 2008, 2009), and powerful
radio halo emission (van Weeren et al., 2009, Bonafede et al.,
2009). We find that the elongated central distribution of galaxies is followed by a very extended tangential critical curve, enclosing a large critically lensed region of $\simeq 2.63 \sq\arcmin$ with an equivalent Einstein radius of $\sim55\pm3.0 \arcsec$ (for an estimated source redshift of $z\sim2.5$). This
corresponds to a physical scale of $350\pm20~kpc/h_{70}$, substantially
larger than in any other known cluster.  This very large radius adds to
the already uncomfortable discrepancy between the large Einstein radii
observed for massive clusters and the predictions based on the
standard $\Lambda$CDM cosmology (Broadhurst \& Barkana, 2008, Sadeh \&
Rephaeli, 2008, Puchwein \& Hilbert, 2009) for which such large
Einstein radii can only be contemplated with mass distributions which
are highly prolate and aligned along the line of sight (Corless \&
King, 2007, Oguri \& Blandford, 2009). Instead, here this cluster is
evidently elongated across the line of sight, traced by the
distribution of member galaxies, and by the extended X-ray emission
and the tangential curves (see Figure 1).

Naturally, such a comparison is biased by the asymmetry and shape of the critical curve, which cannot be perfectly compared with DM simulations. However, this difference is small, as we find that a circularly-symmetric lens of $\theta_E = 55 \arcsec$, centered on the center of Figure 4, contains $6.8 \times 10^{14} M_{\odot}$, 92\% of the mass contained within the tangential critical curve (for $z_s=2.5$).

The mass enclosed within the tangential critical curve, for
$z_{s}=2.5$, is very high $7.4\pm 0.5 \times 10^{14} M_{\odot}$. The
mass enclosed within the larger critical curve for the multiply-lensed
dropout galaxy, number 5, at an estimated source redshift of
$z\sim4$ is correspondingly larger, $\simeq 1.0\times 10^{15}
M_{\odot}$ (marked in red, Figure 1). Clearly, the total mass
associated with this cluster cannot be smaller than this and is likely
to be several times larger, and may soon be reliably estimated from weak
lensing. We stress that the scale of the tangential critical curve hardly
depends on the mass profile because it is set by the critical density
for lensing, which depends only on fundamental constants and the
distances involved, so that different profiles explored in the
modeling procedure yield very similar critical curves, and hence
similarly large Einstein radii and an accurate enclosed mass.

\section{Discussion and Conclusions}

Theoretically, we can make an approximate estimate of the
probability of obtaining such a cluster in the context of the tightly
proscribed $\Lambda CDM$ model. Using the probability distribution function
(PDF) of halo formation times, calculated within the framework of the
extended Press \& Schechter (EPS) formalism (Lacey \& Cole 1993), we
derived a corresponding PDF of halo concentration parameters (Sadeh \& Rephaeli, 2008) by adopting the
formation redshift-concentration scaling, deduced
by Wechsler et al. 2002, from N-body simulations.

 Solving the equation governing the relation between the concentration
 parameter and $\theta_E$ assuming an NFW profile (e.g. Broadhurst \&
 Barkana 2008), leads to a PDF of Einstein radii. Here we adopt the
 mean source redshift $z_s=2.5$ as described above.  The product of
 the Press \& Schechter mass function abundance of clusters within the
 relevant mass and redshift ranges with the cumulative Einstein radius
 probabilities multiplied by the volume available at $z>0.55$ provides
 the number of clusters expected above this redshift for standard $\Lambda CDM$
 where we allow for $2\sigma$ uncertainty in the WMAP5 values of the
 parameters [$n,h,\sigma_8$]). Extremely low numbers of clusters
 are predicted peaking at $M\simeq 4\times 10^{15}M_\odot$ with a maximum
 probability of only $2.7\times 10^{-7}$ objects per Universe (see Figure 5), or approximately
half this value when allowing for the area of sky covered by the ROSAT all-sky survey. This comparison with ideal DM halos is made somewhat uncertain by the observed asymmetry of the central mass distribution. However as noted in section 4, there is only an 8\% difference between the mass within the critical area and the mass within the circular equivalent critical area derived from our model, whereas the discrepancy with theory is orders of magnitude. Clearly it is important to obtain the total mass of the cluster via weak lensing which we are now estimating but it is reasonable to suppose this
lies close to the most probable value derived crudely here, given that
the mass interior to the critical curve is already 25\% of this value.

MACS J0717.5+3745 is the second strong-lensing analyzed cluster
(along the recent critically convergent lens MACS J1149.5+2223, Zitrin
\& Broadhurst 2009) out of several other very luminous X-ray clusters
uncovered by Ebeling et al. (2007) in the ROSAT all-sky survey. Both these clusters have unconcentrated central galaxy
distributions reflecting their unrelaxed state and for which we have
shown that the central mass distribution is likewise very
unconcentrated. A relatively shallow mass profile boosts the
gravitational lens magnification and we calculate that the total area
of sky exceeding a magnification, $\mu>10$, is $\sim 3.5$ square
arcminutes, corresponding to the current high redshift limit of $z\sim
8$, which is far higher than the equivalent area calculated for other
massive clusters (Broadhurst et al., 2005, Zitrin et al., 2009).  Such
unrelaxed and massive clusters opens a potentially new regime of highly
magnifying lenses for accessing the faint distant Universe.

\section*{acknowledgments}
We thank an anonymous referee for useful comments. This research
is being supported by the Israel Science Foundation grant 1400/10.
ACS was developed under NASA contract NAS 5-32865. This research is
based on observations provided in the Hubble Legacy Archive which
is a collaboration between the Space Telescope Science Institute
(STScI/NASA), the Space Telescope European Coordinating Facility
(ST-ECF/ESA) and the Canadian Astronomy Data Centre (CADC/NRC/CSA).


\begin{thebibliography}{}

\bibitem[]{}Bonafede, A., et al., 2009, A\&A, 503, 707
\bibitem[]{}Bouwens, R.J., et al., 2004, ApJ, 616, 79
\bibitem[]{}Brada$\check{c}$, M., et al., 2008, ApJ, 681, 187
\bibitem[]{}Bradley, L.D., et al., 2008, ApJ, 678, 647
\bibitem[]{}Broadhurst, T.J., Taylor, A.N., Peacock, J.A., 1995, ApJ, 438, 49
\bibitem[]{}Broadhurst, T., et al., 2005, ApJ, 621, 53
\bibitem[]{}Broadhurst,T.J. \& Barkana R., 2008, MNRAS,  390, 1647
\bibitem[]{}Broadhurst, T, Umetsu, K, Medezinski, E., Oguri,M., Rephaeli, Y., 2008, ApJ 685, L9
\bibitem[]{}Corless, V.L., King, L.J., 2007, MNRAS, 380, 149
\bibitem[]{}Donnarumma, A., Ettori, S., Meneghetti, M., Moscardini, L., 2009, MNRAS, 398, 438
\bibitem[]{}Ebbels, T.M.D., Le Borgne, J.-F., Pello, R., Ellis, R.S., Kneib, J.-P., Smail, I., Sanahuja, B., 1996, MNRAS, 281, 75
\bibitem[]{}Ebeling, H., Barrett, E., Donovan, D., 2004, ApJ, 609L, 49
\bibitem[]{}Ebeling, H., et al., 2007, ApJ, 661, 33
\bibitem[]{}Franx, M., Illingworth, G.D., Kelson, D.D., van Dokkum, P.G. \& Tran, K., 1997, ApJ, 486, 75
\bibitem[]{}Frye, B. \& Broadhurst, T., 1998, ApJ, 499, 115
\bibitem[]{}Frye, B., Broadhurst, T., Ben\'itez, N., 2002, ApJ, 568, 558
\bibitem[]{}Gavazzi, R., Fort, B., Mellier, Y., Pello, R., Dantel-Fort, M., 2003, A\&A, 403, 11
\bibitem[]{}Gavazzi, R., Adami, C., Durret, F., Cuillandre, J.-C., Ilbert, O., Mazure, A., Pello, R., Ulmer, M.P., 2009, A\&A, 498L, 33
\bibitem[]{}Geller, M. J., Diaferio, A., Kurtz, M.J., 1999, ApJ, 517L, 23
\bibitem[]{}Halkola, A., et al., 2008, A\&A, 481, 65
\bibitem[]{}Kneib, J.-P., Ellis, R.S., Smail, I., Couch, W.J., Sharples, R.M., 1996, ApJ, 471, 643
\bibitem[]{}Kneib, J.-P., Ellis, R.S., Santos, M.R., Richard, J., 2004, ApJ, 607, 697
\bibitem[]{}Lacey C., Cole S., 1993, MNRAS, 262, 627
\bibitem[]{}Liesenborgs, J., de Rijcke, S., Dejonghe, H., Bekaert, P., 2008, MNRAS, 389, 415
\bibitem[]{}Limousin, M., et al., 2008, A\&A, 489, 23
\bibitem[]{}Ma, C.-J., Ebeling, H., Donovan, D., Barrett, E., 2008, ApJ, 684, 160
\bibitem[]{}Ma, C.-J., Ebeling, H., Barrett, E.,,2009, ApJ, 693L, 56
\bibitem[]{}Oguri, M., et al., 2009, ApJ, 699, 1038
\bibitem[]{}Oguri, M., Blandford, R.D., 2009, MNRAS, 392, 930
\bibitem[]{}Puchwein, E. \& Hilbert, S., 2009, MNRAS, 398, 1298
\bibitem[]{}Sadeh, S. \& Rephaeli, Y., 2008, MNRAS, 388, 1759
\bibitem[]{}Springel, V., et al., 2005, Nature, 435, 629
\bibitem[]{}The, L.S., White, S.D.M., 1986, AJ, 92, 1248
\bibitem[]{}Umetsu, K. \& Broadhurst, T., 2008, ApJ, 684, 177
\bibitem[]{}Umetsu, K., et al., 2009, arXiv:0908.0069
\bibitem[]{}van Weeren, R.J., Rottgering, H.J.A., Bruggen, M., Cohen, A., 2009, A\&A, 505, 991
\bibitem[]{}Wechsler R.H., Bullock J.S., Primack J.R., Kravtsov A.V., Dekel A., 2002, ApJ, 568, 52
\bibitem[]{}Zheng, W., et al., 2009, ApJ, 697, 1907
\bibitem[]{}Zitrin, A., et al., 2009, MNRAS, 396, 1985
\bibitem[]{}Zitrin, A. \& Broadhurst, T., 2009, ApJ, 703L, 132

\end{thebibliography}
\end{document}